# Materials for High Temperature Digital Electronics

Dhiren K. Pradhan,[1,a] David C. Moore,[2,a] A. Matt Francis,[3] Jacob Kupernik,[3] W. Joshua Kennedy,[2] Nicholas R. Glavin,[2] Roy H. Olsson III,[1] Deep Jariwala[1,*]

[1]*Department of Electrical and Systems Engineering, University of Pennsylvania, Philadelphia, PA, 19104, USA*

[2]*Materials and Manufacturing Directorate, Air Force Research Laboratory, Wright-Patterson AFB, OH, 45433, USA*

[3]*Ozark Integrated Circuits, Fayetteville, AR, 72701, USA*

**Abstract**

Silicon microelectronics, consisting of complementary metal oxide semiconductor (CMOS) technology, have changed nearly all aspects of human life from communication to transportation, entertainment, and healthcare. Despite the widespread and mainstream use, current silicon-based devices suffer significant reliability issues at temperatures exceeding 125 °C. The emergent technological frontiers of space exploration, geothermal energy harvesting, nuclear energy, unmanned avionic systems, and autonomous driving will rely on control systems, sensors, and communication devices which operate at temperatures as high as 500 °C and beyond. At these extreme temperatures, active (heat exchanger, phase change cooling) or passive (fins and thermal interface materials) cooling strategies add significant mass and complication which is often infeasible. Thus, new material solutions beyond conventional silicon CMOS devices are necessary for high temperature, resilient electronic systems. Accomplishing this will require a united effort to explore development, integration, and ultimately manufacturing of non-silicon-based logic and memory technologies, non-traditional metals for interconnects, and ceramic packaging technology.

[a] Dhiren K. Pradhan and David C. Moore contributed equally to this work.

[*]Authors to whom correspondence should be addressed: dmj@seas.upenn.edu.

## Introduction:

The pervasive demand for embedded electronic systems necessitates development and adoption of electronic computers in non-traditional environments, such as those with high (high-T) or low temperatures, high levels of radiation, or highly corrosive chemicals. In this regard, the application space for high temperature electronics is particularly rich and diverse, requiring new materials solutions for high temperature operation. These include the rapid expansion in avionics and space industries and advanced geo-thermal exploration and nuclear power to meet the growing worldwide energy needs.

The automotive industry has used semiconductor chips operating in the 150 °C temperature range for electronic control of internal combustion engines since the 1980's and more recently in cooling systems of electric motors.[1,2] The recent increase in interest in high-T electronics is mainly driven by industries such as space, supersonic avionics, geothermal exploration and advanced automotives. Further, the advent of "big-data" computing has also necessitated rethinking of the limits and complexity of computing in high-T environments. As a result, mere scaling, reliability and higher operating temperatures for logic devices is no longer the only consideration for advancement. The use of memory devices and performing data-heavy computing operations with limited logic components is equally important and necessary, which has added a new dimension to materials and device research for extreme electronics. Therefore, while numerous reviews exist on individual semiconductors and device classes for high-T electronics, an overarching review that covers all the materials and challenges from semiconductor to system level is notably lacking. Here, we provide a detailed literature review of not only the integrated electronic materials (i.e. semiconductors, metals, dielectrics) and devices (i.e. logic and memory) but also components such as packaging and bonding as summarized in Figure 1.

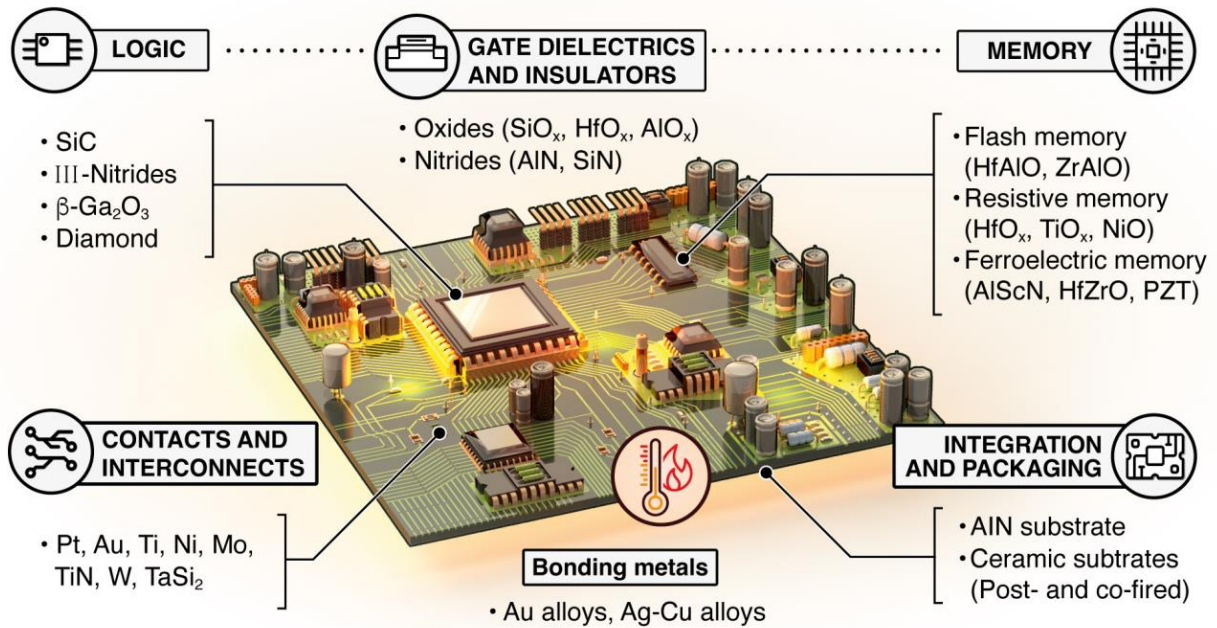

**Figure 1:** Schematic overview figure illustrating the various materials classes critical for high-T digital electronic applications. Selected candidate materials are listed in each class.

# 1. Applications and opportunities for high temperature resilient electronics

Since the late 1970's, the materials, devices, interfaces, and components within silicon integrated circuits have been highly susceptible to degradation and failure at elevated temperatures.[3,4] As a result of these high-T instabilities, the mil-specified limit for operation of most silicon devices is typically restricted to below 125-175 °C, although many of the electronic components may operate at higher temperatures for a short period of time (Box 1).[5] In ambient environments, overheating of device components are typically mitigated by passive or active cooling strategies,[6,7] but these can be impractical in conditions where the surrounding environmental temperature is elevated resulting in failure of silicon devices (Box 2). To combat the exceptionally high junction leakage current experienced in silicon devices at elevated temperatures, silicon-on-insulator (SOI) technology has been employed for device operation up to about 300 °C.[8] This is accomplished via introduction of a dielectric layer on top of the bulk CMOS substrate underneath the semiconductor can reduce leakage currents by as much as three orders of magnitude at 250 °C.[9,10] Moreover, the output conductance has actually been shown to improve as temperature is increased in SOI architectures.[11] SOI Metal Oxide Semiconductor Field effect Transistors (MOSFETs) have demonstrated operating at temperatures even approaching 450 °C,[10] albeit for a short period of time with reduced functionality, and have even expanded into use cases for high temperature flexible[12] and RF[13] devices. In general, however, SOI technology is fairly restricted to temperatures below 300 °C, with new materials solutions required for higher temperatures. The use cases for electronics operating > 300 °C is summarized in Box 1 and Table 1 below.

Most of the above high-T electronic applications have been analog in nature. The clear need for digital electronics for high-T environments stems for the need for more sensors which in turn leads to the computing and feedback control in-situ which is unfeasible with analog electronics. Further, interpretation of collected data requires use of neural network architectures, which, while easy to implement in software, may not have the hardware for implementation at elevated temperatures. Keeping this in mind, scaling and performance enhancement of digital logic combined with use of memory (both volatile and non-volatile) is critical. These aspects of digital electronics hardware for high-T operation are highlighted and discussed below in detail.

---

**Box 1: Use cases for High-T (>300 °C) electronics**

There have long been energy-centric, critical applications in "harsh environments," or environmental conditions strongly deviating from room-T and ambient pressures/atmosphere. The first use-cases for high-T electronics were in the exploration of energy from chemical resources and power generation from thermal resources. Drilling produces heat, and drilling of wells in ever more diverse geological formations has required electronics to steer the drill, log the well, and increase the likelihood of a successful field development.[14] This led to the Department of Energy's Deep Trek program in the early 2000's to help develop SOI technologies that could operate above 200 °C for these purposes.[15] As energy technology moves beyond oil and natural gas, extraction of energy from geothermal reserves have the potential to power the entire planet; but accessing this resource requires drilling new geothermal wells, 10 miles and deeper.[16,17] This will require electronics to operate at intense temperatures of 400 °C or more. Further, directional drilling and advanced logging/monitoring are critical, since in geothermal wells the surface area for heat transfer must be maximized, either by finding natural porous formations, by drilling large lateral fields, or by opening (fracturing = fracking) rock to enable heat to transfer.[18] Likewise, generation of power from nuclear sources also requires operating in the presence of radiation and monitoring heat transfer media in emerging molten salt reactors.[19,20] In each of these applications, the electronic devices and sensors experience both high static temperatures and high thermal ramp rates.

High temperatures have always been a byproduct of converting energy into motion for transportation. Automotive electronics have long been a driver of harsh environment electronics.[1,14] Electronic control was one of the great revolutions of automotive combustion engines, beginning largely in the 1980s. From 2004-2016, advanced electronic controls allowed fuel economy to improve by 32% in the US.[21] With engines operating as high as 2500 °C in the center of the combustion chamber, temperature ranges experienced by sensors and electronics have been extended to 150 °C, with specifications such as AECQ100.[22] Electric cars have complex cooling systems in part to manage the heat generated by conversion of direct current battery energy to three phase electronic drives, with 14%–33% of the total volume of traction inverters dedicated to cooling.[2] Higher temperature power and control electronics can work to remove this burden and increase efficiency, in a sector that generates 14% of $CO_2$ emissions.

In aircraft, temperatures from ambient to more than 1700 °C in the hot section represent a significant challenge for distributing sensors and electronics across various platforms. Aircraft engines, similar to automotive engines, have been early adopters of electronic control.[23-25] Today, all modern turbine engines use digital control with advanced sensors, which lowers maintenance cost, reduces weight, and increases efficiency.[26] Full Authority Digital Engine Controls in use today are thermally managed (typically to <125 °C by part availability), reaching the limit of their usefulness and ability to control larger and more complex high-bypass designs. As fuel is the primary 'coolant' in turbines, electrification and alternative fuels like hydrogen mean higher temperature electronics are needed.[27,28] Advanced hybrid turbines supporting sub, super and potentially hypersonic flight will need higher temperatures to operate, integrating higher power level electronics in a thermal limited environment; in a sector that contributes over 2% of global $CO_2$ emissions.[29]

| Phase Transitions / Temps of Interest | °C | Standard Temp. Range [Organic Packaging] (Commercial, Industrial, Automotive, Military) | Extreme Temp. Range [Ceramic Packaging] (SOI CMOS, WBG, UWBG) | Environments | Substrates |
|---|---|---|---|---|---|
|  | 1000 |  |  |  |  |
|  | 800 |  |  | Hypersonic Flight |  |
| Magma | 700 |  |  | Molten Salt Reactors | Postfire Ceramic Systems |
|  | 600 |  |  |  |  |
|  | 500 |  |  |  |  |
|  | 400 |  |  | Enhanced Geothermal |  |
|  | 362 |  |  | Venus surface |  |
| Lead melts | 327 |  |  |  |  |
|  | 315 |  |  | Internal Combustion Engines | Cofired Ceramics Systems |
|  | 300 |  |  | Down Hole Oil Exploration |  |
|  | 225 |  |  |  |  |
|  | 150 |  |  | Automotive (high) |  |
|  | 125 |  |  | Military Aerosapace (high) |  |
| Liquid water boils | 100 |  |  | Industrial Electronics (high) | Organics |
|  | 85 |  |  | Commercial Electronics (high) |  |
|  | 15 |  |  | Mars surface, daytime (high) |  |

**Table 1.** Summary of use cases and high temperature ranges for electronics and corresponding packaging materials. ( Grey and green colors indicate standard temperature and physically possible operation respectively. The red and grey boxes in the "Environments" column indicate research and commercial applications, respectively)

> **Box 2: High temperature limits of silicon electronics**
>
> The high temperature failure mechanisms in silicon electronics vary from intrinsic material limitations (i.e. reduction in carrier density and dielectric breakdown) to reliability issues and integration challenges (i.e. electromigration in contacts and failure of metal junctions).[30] The intrinsic carrier density, strongly dependent upon the bandgap of the semiconductor, is one metric that defines an ultimate temperature limit for practical device operation. As there is an exponential dependence relating intrinsic carrier density to temperature, the low bandgap of silicon relative to other semiconductors such as SiC, GaN, Diamond, and $Ga_2O_3$ restricts the absolute limit for silicon device operation to roughly around 400 °C.[30] At even lower temperatures, other effects including increased leakage current above 0.1 A/cm$^2$ further reduces operation limits near 250 °C.[31] In the silicon dioxide ($SiO_2$) dielectric, increased leakage current, accelerated time-dependent dielectric breakdown, and hot carrier degradation are all accelerated at elevated temperatures above 250 °C.[30] Other considerations include electromigration of metal interconnects such as aluminum at temperatures of 200-300 °C, Sn-Ag-Cu (SAC) and SnSb alloy solders can melt at temperatures around 200-240 °C, and generation of compounds between Al/Au metals at wire bonds can restrict operation above 400 °C.[32,33]

## 2. Emergent logic materials and devices beyond silicon

### a. Wide bandgap semiconductor materials - GaN, SiC, Diamond

Logic devices are at the heart of any modern digital electronic system. In this section, we will highlight materials innovations in logic devices that allow operating beyond Silicon limits, in WBG and UWBG semiconductors. While there are several WBG and UWBG semiconductor candidates, few have been thoroughly studied and experimented with for high-T logic applications. Among them, SiC, III-nitrides (GaN, AlGaN, AlN) and diamond have been the most heavily investigated. Some oxides, namely β-$Ga_2O_3$ and $Al_xGa_{1-x}O_3$, have also been investigated. It is worth noting that simply relying on band gap size at room T is not sufficient for selecting a semiconductor for high T logic applications. Instead, the dependence of intrinsic carrier density vs. T is more important since it accounts for temperature induced changes in band-gap as shown in Figure 2 a. 4H-SiC has a unique advantage in this regard compared to most other WBG semiconductors particularly wurtzite GaN and β-$Ga_2O_3$ at temperatures > 850 °C given the minimal dependence of its band gap on temperature.[34-39]

### i. Silicon Carbide (SiC):

For temperatures > 300 °C, SiC is the most mature semiconductor technology with complementary p and n doping readily available and multiple logic device types available and investigated. The main reason for this is the size of the band-gap of the now common polytype, 4H, (3.4 eV) and availability of up to 8" (200 mm) 4H wafer substrates. This makes mass production of complex integrated circuitry possible adopting some of the same principles that have been perfected for Silicon over the decades.[40-42] The advanced level of maturity of SiC as a high-T semiconductor material is largely a result of its development for both power and high-T electronics since the 80's. High-T SiC MOSFETs operating at 650 °C were demonstrated as early as 1987.[43] . The advent of commercial 6H (1991) and 4H (1994) SiC wafers from Cree Research truly jump-started SiC electronics and integrated circuits research for high-T electronics. Thorough reviews of the early works in this area are available.[44,45] We will therefore focus on more recent advances in this field. A major development in SiC very high-T logic technology over the past decade has been the advent of Junction FETs (JFETs). In particular, the NASA Glenn center group have shown through a series of results that normally-on n-channel SiC JFETs can have stable high temperature operations for thousands of hours at ~500 °C under ambient atmosphere[46-49], while shorter term (<150 hours) demonstration of packaged logic circuits up to 961 °C under ambient have been reliably made.[50]

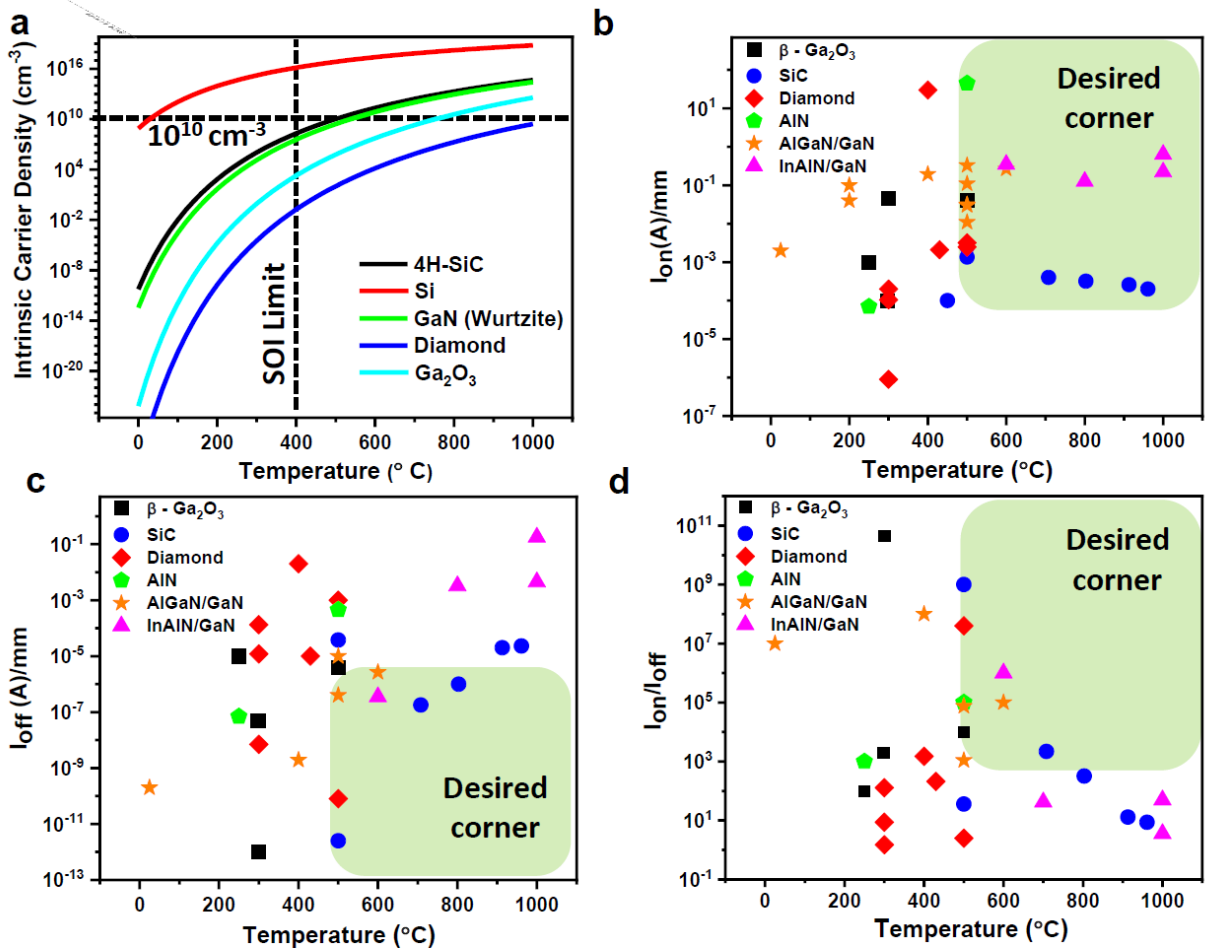

**Figure 2. Wide Band Gap (WBG) Semiconductor Materials and Logic Devices for High Temperature Operation.** a. Intrinsic carrier density vs. T for Si and prominent wide-band gap semiconductor materials.[14,30,34,37,51,52]. The low variation of 4H-SiC band gap with increasing temperature allows it to maintain lower intrinsic carrier density as compared to both wurtzite GaN and β-$Ga_2O_3$ at temperatures above 850 C. b-d. ON state current density, Off state current density and ON/OFF current ratios for notable reports of high T logic devices.[42,46-50,53-82] Desired performance corners are indicated.

This technology platform includes demonstration of inverters, transistor + resistor logic gates, D-flip flops, ring oscillator clocks and volatile memory like SRAM to perform full scale digital computing tested at 500 °C for up to a year.[48] This technology was upscaled in 2021[56] and further in 2023[42] with a larger die size of 55 mm and reduced feature dimensions including development of a robust back-end-of-line (BEOL) interconnect process. However, 500 °C long-term (~ year long) cycling of the interconnects and the upscaled chips with high yields remains to be achieved. These demonstrations suggest that JFET IC technology with appropriate packaging and protection can truly become the dominant hardware technology for high-T computing. Conversely, SiC MOSFETs have also been investigated since the early days, however, they face fundamental challenges pertaining to charge traps at the $SiO_2$/SiC (gate dielectric/semiconductor) interface. In particular, the $SiO_2$/SiC interface has significant density of interface trap states which are close to the conduction band edge but have a significantly long tail below the conduction band edge. As the temperature rises, the occupancy of these trap states reduces since the Fermi level moves further below the conduction band edge due to thermal generation.[83] Consequently, there is a significant, negative threshold voltage shift for n-MOSFETs and vice versa for p-MOSFETs. Simultaneously, the field effect mobility can increase at low-overdrive voltages with increasing temperatures due to the reduction in interface trap occupancy, which contributes to carrier scattering at the interface. This effect reduces at high overdrive voltages due to higher phonon scattering at higher temperatures, which reduces bulk mobility.[83] The above effects make MOSFETs much less reliable for high-T operation and repeated thermal cycling. Nonetheless, much progress has been made with high-T SiC MOSFETs. SiC MOSFET operation has been demonstrated for over 100 hours at 470 °C and inverters operating upto 200 °C.[59] Ring-oscillators and other digital logic circuits have also been tested at 470 °C (Venus surface temperatures). [84]Three major advances include: i. Using trench-type (fin-like) channels instead of planar channels which helps flatten the density of interface states and reduces sensitivity of threshold voltage to temperature.[85] ii. using nitrogen (nitric oxide) assisted oxidation as well as post oxidation annealing in $POCl_3$ to incorporate P atoms into the oxide, both of which have helped reduce the interface state density.[86] iii. Using tri-layer ($SiO_2$/SiNx/$Al_2O_3$) dielectric stacks combined with a ring channel structure where the SiNx serves as a reaction barrier layer and prevents migration of Al ions at high T, while $Al_2O_3$ prevents electron injection from the gate into the channel.[57]

While SiC logic is the most mature and advanced at elevated temperatures (Figure 2 b-d), it has its own set of challenges including difficulty in fabricating ohmic contacts for p-type FETs.[87]. None the less, it is the only high-T logic technology where stable operation of complex integrated circuits at elevated temperatures for up to a year of operation has been demonstrated which makes it the most suitable near-term candidate for scalable and complex high-T microprocessors.[88]

**ii. III-Nitrides**:

III-Nitrides which include GaN, AlGaN and AlN have recently become the most heavily-investigated semiconductors for high-T logic materials and devices. There are several positive attributes to the nitrides, the most important ones being the larger band-gaps and higher peak and saturation drift velocities for the charge carriers when compared to SiC. Further, the higher thermal conductivity compared to Silicon is also advantageous. The ability to epitaxially modulate compositions in the ternary $Al_xGa_{1-x}N$ ($0 \leq x \leq 1$) system to form 2-dimensional electron gasses (2DEGs) makes this material class attractive for high-T high-electron-mobility transistors (HEMT) devices. A tremendous amount of work has been reported on nitride logic devices for both high-T and high-power devices. Here we will again focus on the most recent results relevant to high T operation and direct the reader to other reviews that comprehensively cover prior works.[89-91] HEMT devices are the dominant class of nitride logic for high T operation. Among HEMTs, there are 2 varieties. The first variety is the JFET type with a p-GaN contact on a AlGaN (barrier)/GaN whose interface supports the 2DEG.[67] Some versions of these transistors are often utilized in Direct Coupled Field-Effect Transistor Logic (DCFL). The second variety is the metal–insulator–semiconductor field-effect transistor (MISFET) type that have an insulator such as $Al_2O_3$ on top of the AlGaN through which the gate electric field is applied.[68] In both types of devices, enhancement (E-mode) and depletion (D-mode) devices have been demonstrated.[68,70] These two modes (D/E) when combined can be operated such that the D-mode device serves as the driver while the E-mode device serves as the load leading to operational inverters and ring oscillators at elevated temperatures up to 500 °C[70] as well as SRAM up to 300 °C.[92] A notable achievement for the E-Mode JFET type HEMTs is testing at elevated temperatures and pressures (460 °C, 92 atm., simulated Venus environment) for extended periods (~10 days) and achieving ~18 % current degradation and 0.02V in threshold shift.[71] Similar to the case of SiC channels, the JFET variety have exhibited more thermal hardness as opposed to the MISFET variety for GaN HEMTs. However, use of multilayer dielectrics and a circular (ring shaped) gate/channel geometry have shown improved thermal hardness for MISFET type GaN HEMTs.[73] Overall, GaN HEMT technology is the second most mature behind the SiC JFET. However, this technology also suffers from lack of complementary devices. In addition, the degradation of the 2DEG mobility and hence drive current due to phonon scattering at elevated temperatures is well documented.[69] This is a primary drawback of GaN based transistors as compared to SiC based JFET technology. A very promising strategy in terms of GaN based HEMTs to achieve comparable performance to SiC based JFETs is the use of InAlN (In < 20%) as the barrier layer instead of AlGaN. InAlN has shown to be extremely well lattice matched to GaN and allows for concurrently high sheet carrier density ($>10^{13}$ cm$^{-2}$) together with low to absent mechanical stress in the heterojunction.[93,94] The device structure typically comprises of an ultrathin (< 10 nm) thick InAlN barrier layer on a GaN layer (varying thickness) with a ~1 nm AlN seed layer. This gives InAlN/GaN HEMTs with exceptional thermal stability with early demonstrations of operation made upto 1000 C in vacuum for much degraded performance[94] followed by 900 C stable operation for 50 hours.[93] Later studies showed ON current density ~0.4 A/mm combined with an ON/OFF ratio of $10^6$ at 1 MHz at 600 C[95] which remains among the best overall performance for a nitride transistor at high-temperatures.

Among the nitride class of materials, the push has always been to even higher band gaps. AlGaN HEMTs and metal-semiconductor field-effect transistors (MESFETs) have also been researched and demonstrated at elevated temperatures. In this case pure AlN replaces AlGaN as the barrier layer whereas the 2DEG is formed at the AlN/AlGaN interface.[66,75] Such devices show impressive ON/OFF ratios (~$10^7$) and sub-threshold swing (~75 mV/dec). However, the carrier mobilities shown have been limited (~250 cm$^2$/V.s) and high temperature demonstration has also been limited to 125 °C.[75] Additionally, minimal ON current degradation in similar devices up to 300 °C has been demonstrated.[66] AlGaN (n-i) homojunction MESFETs grown on high quality AlN on sapphire have shown to be stable up to 200 °C with no noticeable degradation in ON current.[74]

The largest bandgap in the III-nitrides is found in pure AlN. AlN has been a subject of interest for deep UV opto-electronics as well as high-T and high-power electronics for a very long time.[96-98] However, growth of high quality thin-films and complementary doping have posed some of the greatest barriers for AlN in terms of device demonstrations. Recently, there has been significant progress in epitaxial growth and complementary doping of AlN.[97] In particular, achieving p-doping of AlN via Be dopants at a concentration > $1 \times 10^{18}$/cm$^3$ using metal modulated epitaxy (MME) has been the key breakthrough.[99] Still, very limited reports exist on AlN channel transistors operating at elevated temperatures.[64,66] Unipolar n-MESFETs operating at 500 °C with epitaxial AlN grown on SiC have none the less been recently demonstrated with ON/OFF > $10^5$ and ON current density > 0.01 A/mm at 500 °C.[65]

### iii. III-Oxides:

Similar to nitrides, Ga and Ga-Al oxides have also been proposed and experimented as UWBG semiconductors for high temperature applications. Among them β-$Ga_2O_3$ is the most heavily investigated phase due to its stability at high temperatures and is therefore the most mature in terms of technology development with 4" substrates available commercially.[100] While the nitrides are more mature with several works available across a range of compositions, as in the early days of SiC, work in oxides has been focused on high power devices taking advantage of its high breakdown field strength. None the less, there has been substantial interest in β-$Ga_2O_3$ field-effect devices with applicability to high temperature, with the first transistor demonstration in 2012.[101] More recently, β-$Ga_2O_3$ MOSFETs on AlN substrates with $Al_2O_3$ dielectrics have been made and measured up to 300 °C with ~30 % degradation of ON current at 300 °C compared to room T.[63] β-$Ga_2O_3$ power MOSFETs have also been tested at elevated temperatures in multiple reports.[102,103] A similar degradation in On current and a pronounced reduction in ON/OFF ratio at 300 °C is observed due to rising OFF currents.[103] In other recent work, Fin-FET structure devices have been demonstrated using epitaxially grown Si doped β-$Ga_2O_3$ on semi-insulating β-$Ga_2O_3$ and have been compared with planar MOSFETs for high T operation up to 300 °C. Most temperature dependent device characteristics between Fin vs. planar FETs were observed to be comparable suggesting no distinct advantage of the Fin geometry[62], unlike in the SiC case. However, this area remains open for future research. Since MOSFETs present significant challenges at high T due to the imperfect semiconductor/gate-dielectric interfaces, MESFETs (similar to JFETs) have also been investigated for β-$Ga_2O_3$ tested up to 500 °C. While they show lower ON/OFF ratios closer to room T, they show minimal degradation of ON current at 500 °C and even show recovery upon soaking for 1 hour at 500 °C suggesting thermal annealing improves electrical characteristics of β-$Ga_2O_3$.[61]

### iv. Diamond:

Diamond as a high temperature (and high voltage) UWBG semiconductor material has fascinated material scientists and engineers alike for decades, with the first bipolar transistor action in natural diamonds demonstrated in 1982[104] and first high T point contact transistors operating at > 500 °C demonstrated in 1987 on synthetic diamonds.[105] The first thin-film, diamond high T (~300 °C) FETs were demonstrated in the early 90's, albeit with poor channel modulation.[77,106] Most Diamond FETs studied have been p-doped in the bulk with Boron (B) and hence show n-channel behavior. In recent years, the use of hydrogen terminated diamond surfaces as two-dimensional hole gases (2DHG) have also been prevalent for both high frequency and high temperature applications. In addition, bulk n-doping of diamond via Phosphorus (P) has been accomplished.[107] Among the most prominent results is the demonstration of JFETs with a specific on-resistance < 2 mΩ·$cm^2$ over 400 °C, leakage currents of $10^{-15}$–$10^{-13}$ A, and ON/OFF ratios > $10^6$ at 450 °C.[78] Among MOSFETs, 400 °C operation with ON/OFF ratios ~$10^6$ in p-channel devices has been achieved by Silicon passivation of diamond surfaces with $SiO_2$ gate dielectrics to stabilize 2DHG.[108] In contrast, inversion p-channel MOSFETs with alumina dielectrics observed irreversible degradation upon heating to 400 °C attributed to threshold voltage shift from a reduction of the total density of the extrinsic charges at the $Al_2O_3$/diamond interface with unintentional post-deposition annealing of the $Al_2O_3$ gate oxide at high temperature.[109] Despite much progress in diamond FETs over the decades, integrated p and n devices remain challenging and studies on long duration temperature hardness testing remain unavailable. Some recent works have explored hetero-integration of n-channel GaN and p-channel diamond MOSFETs for high T inverters but their performance remains far from SiC JFETs, even at 250 °C.[110]

**v. Vacuum channel devices:**

Vacuum channels have always been attractive for temperature and radiation hard applications because of the high electron velocity in vacuum. In recent years, the sophistry of nanoscale fabrication has brought alive the field of miniaturized vacuum devices once again, achieving channel dimensions < 50 nm, which is the mean free path of air molecules at atmospheric pressure.[111] This means highly scaled vacuum channel devices do not necessarily need to be packaged or operated in high vacuum. The demonstration was made by a NASA Ames lab team using sharpened needle like source and drain electrodes laterally etched in a Si wafer and a doped polysilicon gate with a high-k $HfO_2$ dielectric hosting a cylindrical vacuum channel (similar to a gate-all-around structure without any semiconductor). These devices made on an 8" Si wafer show ON/OFF ratios > 1000 and 3 µA drive current at 2 V. Temperature hardness up to 200 °C and radiation hardness to proton and gamma ray doses were also demonstrated.[112] However, for high-T operation, SiC instead of Si is a better material choice and a vertical instead of a lateral device geometry is favorable. This is demonstrated in a recent work from the same NASA team over 150 mm SiC wafers though no elevated temperature testing was performed.[113]

**a. Dielectric insulator materials**

In high temperature applications, the dielectric materials in the BEOL have several important functions and properties. In low temperature electronics, a low dielectric permittivity is of the highest importance as the switching speed of the integrated circuits is often limited by the BEOL capacitance. While the dielectric permittivity plays the same role on high frequency switching in high temperature electronics, other considerations such as the coefficient of thermal expansion (CTE) mismatch with the substrate[114] and other layers, the ability of the dielectric to protect the backend metallization from oxidation,[48] and the surface and bulk resistivity of the dielectrics at elevated temperature are also of significant importance.[114] Similar to low temperature electronics, the BEOL dielectric materials should also be amenable to a conformal deposition method capable of depositing relatively thick films (~1000 nm), such as chemical vapor deposition (CVD), in order to ensure conformal coverage of the BEOL metallization.

An excellent example of the complex tradeoffs involved in high temperature BEOL dielectric stack design is the SiC JFET – resistor logic developed by NASA Glenn that has demonstrated operation over 1 year at 500 °C[48] in air ambient. The technology has a 2-level BEOL $TaSi_2$ metallization with three dielectric layers. The dielectric between the SiC and metal1 (M1) is 30 nm of thermally grown $SiO_2$ plus 1000 nm of low-pressure chemical vapor deposited (LPCVD) $SiO_2$. A 1 µm thick LPCVD $SiO_2$ forms the dielectric between M1 and metal2 (M2). Encapsulating the chip is a top dielectric stack of $SiO_2$ (1000 nm)/$Si_3N_4$ (67nm)/ $SiO_2$ (1000 nm). This encapsulation stack successfully prevents oxidation of the $TaSi_2$ metallization at high temperature even in an atmospheric air environment while the $Si_3N_4$ layer serves the additional purpose of mitigating mobile ion contamination.[54]

While the NASA SiC JFET–resistor logic process has demonstrated the longest operation at extreme temperatures, the operational lifetime at different elevated temperatures and over thermal cycling has been shown to be limited by cracking of the BEOL dielectrics, leading to subsequent oxidation or breaking of the BEOL metallization and in some cases peeling of the dielectrics.[115] While the dielectric stack outlined above included a $Si_3N_4$ layer embedded only in the top $SiO_2$ dielectric layer,[48] a study of 6 different BEOL dielectric stacks,[42] embedding 100 nm of $Si_3N_4$ between one, two, or three of the BEOL $SiO_2$ layers, showed significant differences in cracking during the final high temperature $SiO_2$ deposition. These studies highlight the primary importance of CTE mismatch, film stress, and oxygen permeability when choosing BEOL dielectric materials and material stacks for high temperature electronics.

### 3. Emergent memory materials and devices

While high-T logic is relatively mature, and has been under development for decades, non-volatile memory remains an Achilles heel for high-T digital computing. Volatile memory, in addition to not being able to retain state, is also expensive (in terms of space and transistors) as well a power consumption (high-T leakage). Therefore, memory limits scaling of computing in elevated temperature environments and development of stable, compact, low-power NVM for elevated temperatures is a pressing research challenge. Many NVM technologies use a change in resistance of the active material as the logical state. Phase change memory (PCRAM) relies on crystalline amorphous phase transition in germanium tellurides where recrystallization occurs at ~200 °C while magnetic memories such as magnetic tunnel junction (MTJRAM), and spin torque transfer (STTRAM) have obvious temperature limitations due to Curie temperatures (~500 °C) of the ferromagnetic alloys of nickel, iron, and chromium.[116,117] However, for practical purposes the robustness of the remnant magnetization in these structures is reduced in favor of low switching power by employing very thin magnetic conductive layers. This means that commercial magnetic RAM is typically limited to operating temperatures less than 125 °C.[118] Here, we will focus on flash, ferroelectric and resistive memory three most prominent NVM technologies for high-T operation. A concise summary of these NVM technologies and their performance vs temperature is provided in Figure 3.

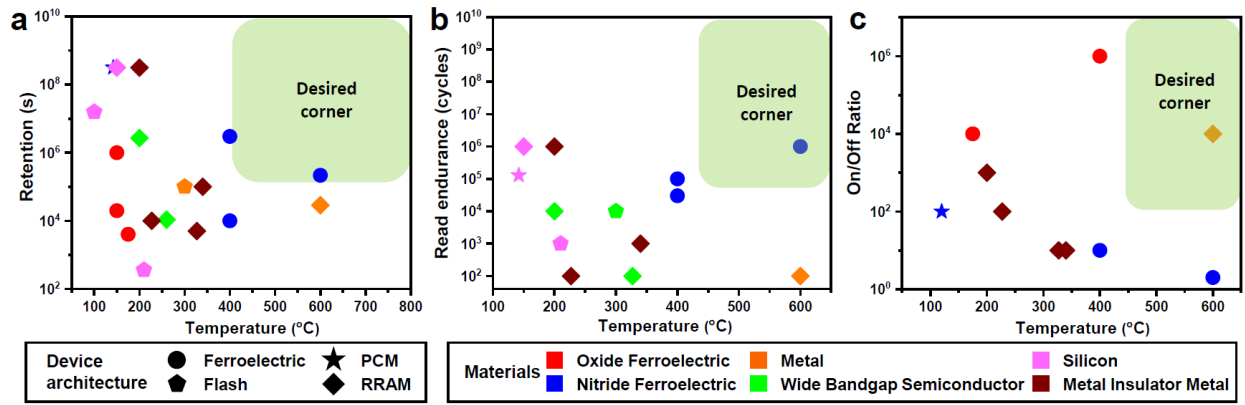

**Figure 3. NVM memory for high-T digital electronics. Shapes designate device type, and colors designate materials** a. Data retention vs. temperature. b. Read endurance vs. temperature. c. ON/OFF ratio vs temperature for various demonstrated NVM technologies. Nitride Ferroelectric NVM appears the most promising technology. However, much progress remains to be achieved to reach desired corners on all three figures of merit.[119-139]

#### a. Flash memory:

Flash memory is primarily not developed for extreme environments, but it is state-of-the-art for most commercial electronics operating at high-T. Flash memory is a well-developed technology, and is the standard for commercial NVM in most consumer electronics and computing. First developed in 1980 at Toshiba, and first sold in 1987,[140] technological improvements through the years have primarily focused on speed and capacity rather than robust temperature endurance because consumer electronics rarely face extreme environments. Commercially, flash memory exists in two types, NAND and NOR flash, both of which use identical architectures and doped silicon MOSFET materials. The tradeoffs of flash memory architectures is well covered;[141,142] for the purposes of this survey consider that NAND flash can generally be constructed more densely and cheaply, whereas NOR flash can be more reliable, though advances in NAND reliability have allowed it to dominate the commercial market.[143,144] Therefore, most large memory applications, such as solid-state drives and USB drives use NAND arrays, whereas NOR flash arrays are more common in device controls. As discussed below, some FET-based high temperature devices have replicated flash architectures successfully, usually opting for NOR-type's reliability, but these were not designed for extreme environments.

Temperature concerns in commercial flash memory are focused on the problem of thermal management; memory operation creates heat that needs to be shed to maintain performance and reliability. Generally, consumer memory begins to lose performance around 85 °C, begins to degrade around 150 °C, and rapidly degrades at 210 °C (material details in Box 2).[120,121,145,146] These temperatures are acceptable for most computing applications, but not in extreme environments.

Material improvements can also improve temperature performance, both by reducing performance drop prior to failure and by increasing failure temperature. Notably, wide-bandgap materials such as GaN, GaP, 4H-SiC, and 6H-SiC have been used to construct MOSFET or JFET transistors which can be configured into operable NVM. These have been successfully tested as operable at 300 °C,[92] 275 °C,[147] 500 °C,[48] and 452 °C[49], respectively. Another approach that yields high-temperature survivability is self-annealing configurations via heating dispersal,[148,149] which have achieved single-transistor repair at 800 °C, but have not been demonstrated to operate in extreme environments.

**b. Ferroelectric memory**

Ferroelectric (FE) materials exhibiting high ferroelectric Curie temperatures ($T_C$) have strong potential for utilization in NVM devices operating at very high temperatures (> 500 °C). Ferroelectric materials exhibit switchable and spontaneous polarization, electric field-driven switching and an ability to maintain the switched polarization state for long time. These properties make them suitable for low-power nonvolatile memory applications as they exhibit fast switching speed, lower switching energy, and have potential for multibit operation.[150-152,153] FE materials can be configured into memory devices in many ways, the most prominent of which are ferroelectric RAM (FeRAM), Ferroelectric Field Effect Transistors (FeFETs), Ferroelectric tunnel junction (FTJ) memristors, Ferroelectric diode (FeD) memristors, and Ferroelectric domain wall memory (FDWM).[154,155] To meet the temperature demands of harsh environment applications, FE materials with high Curie temperatures are required because polarization switching and stable retention cannot be observed if the device temperature is close to $T_C$.[156]

Ferroelectric memory technologies utilizing perovskite and/or fluorite structured oxides such as $Pb(Zr_{1-x}Ti_x)O_3$ (PZT), $BiFeO_3$ and Hf-based oxides have been investigated, but these materials face temperature limitations.[157-160] Among the ferroelectric perovskite oxides, PZT is one of the best, having a maximum remanent polarization ($P_R$) of ~ 70 µC/cm$^2$ with a ferroelectric $T_C$ less than 500 °C.[161,162] Samsung achieved a milestone in 1996 by manufacturing a 4 MB FeRAM utilizing PZT grown by chemical solution deposition. The utilization of PZT in FE memory technology still has many issues, such as it is not fully compatible with CMOS technology and

BEOL integration because of high crystallization temperatures, fast diffusion of Pb in Si, and weakly bound oxygen.[161] In addition, PZT suffers from the destabilization of the polar structure, enhanced domain wall mobility, and chemical instabilities due to the volatile nature of lead at higher temperature.[159-161] Among the other perovskite ferroelectrics, bismuth iron oxide ($BiFeO_3$) is one of the mostly studied lead-free ferroelectric materials having a high $T_C$ of ~ 830 °C.[161,163,164] However, this material has several drawbacks such as high leakage, poor switching speed, and low cycling endurance compared to other well-known ferroelectrics.

Another hexagonal structured perovskite ferroelectric material, $LiNbO_3$ exhibits a very high ferroelectric $T_C$ of ~ 1210 °C.[165] A ferroelectric nonvolatile domain wall random access memory was fabricated utilizing $LiNbO_3$ which can be operated up to 175 °C.[133] $SrBi_2Ta_2O_9$(SBT), another bi-layered Aurivillius structured ferroelectric had gained significant attention as Panasonic demonstrated 4 MB FeRAM using this material. SBT exhibits a ferroelectric $T_C$ of ~ 300 °C, $P_R$ ~ 10-15 $\mu C/cm^2$ and coercive field ($E_C$) < 0.1 MV/cm.[154,166,167] The lower $T_C$, $P_R$ and $E_C$ along with the contamination due to mobile element Bi in SBT are the major limitations to be utilized high temperature NVM devices.

Several materials with perovskite-like layered structures have been demonstrated to exhibit ferroelectric $T_C$ above 1000 °C. Among the notable examples are $Sr_2Nb_2O_7$, $Ca_2Nb_2O_7$, $La_2Ti_2O_7$, $Pr_2Ti_2O_7$, and $Nd_2Ti_2O_7$ exhibit ferroelectric $T_C$ ~1327, >1525, ~ 1500, > 1560, and 1450 °C respectively. The $P_R$ < 10 $\mu C/cm^2$ and the $E_C$ < 0.1 MV/cm were observed for all the perovskite like layered structured materials. The low $P_R$ and $E_C$ values of these materials might hinder them for integration into high-T NVM applications but warrant comprehensive research efforts.[168-171]

In 2006, a significant breakthrough occurred with the discovery of ferroelectric behavior in silicon (Si)-doped $HfO_2$ featuring a fluorite structure, which has opened up exciting technological possibilities, enabling the seamless integration of ferroelectric materials into 28 nm and 22 nm Si CMOS nodes.[172-175] Despite the promising outcomes exhibited by $HfO_2$-based ferroelectric materials its polymorphism crystal structures introduces processing challenges. Effectively suppressing the stable monoclinic phase while enhancing the metastable ferroelectric phases poses a hurdle. This situation can result in significant device-to-device discrepancies, potentially raising concerns about reliability, particularly in the context of employing it for large-scale integrated circuits and at high temperature. Despite the potential for favorable cycling endurance and memory state retention in doped $HfO_2$, a notable decrease in $P_R$ above 623 K (~ 350 °C) was observed in a mixed zirconium hafnium oxide ($Hf_{1-x}Zr_xO_2$) system.[176] In this study, it was also reported that the ferroelectric $T_C$ is dependent on $ZrO_2$ content, but it does not exceed 800 K (~ 527 °C) irrespective of the composition.[176] In addition, doped hafnium oxide based materials exhibit $P_R$ of 10–50 $\mu C/cm^2$ and $E_C$ of ~ 1–1.5 MV/cm.[154] In another report, the ferroelectric $T_C$ of HZO is predicted to be above 1000 °C, and HZO/$\beta$-$Ga_2O_3$ ferroelectric FET devices can be operated up to 400 °C, but this device shows a sharp decline in polarization at 300 °C.[134]

The groundbreaking discovery of ferroelectricity in a new class of nitride materials: wurtzite-structured aluminum scandium nitride ($Al_{1-x}Sc_xN$) having a hexagonal crystal structure was made in 2019.[177] Since 2009, it has been well-known as a good piezoelectric material utilized in several electronic components. Ferroelectricity has now been demonstrated in numerous additional wurtzite-structured III nitrides such as $Al_{1-x}B_xN$, $Ga_{1-x}Sc_xN$, $Al_{1-x}Y_xN$ and this phenomenon extended to wurtzite-structured oxides such as in $Zn_{1-x}Mg_xO$.[178-183] Interestingly, all these wurtzite-structured nitrides and oxides also exhibit high $P_R$ (> 100 $\mu C/cm^2$) and $E_C$ > 2 MV/cm. Except for $Al_{1-x}Sc_xN$ and $Al_{1-x}B_xN$ the investigation of temperature-dependent physical properties and the exploration of the ferroelectric $T_C$ are still in the early stages of research.[119,135,184] $Al_{1-x}Sc_xN$ is the mostly studied material among the wurtzite-structured nitrides having high $P_R$ (80-140 $\mu C/cm^2$), tunable $E_C$ > 2-6 MV/cm, bandgap (4 –5.6 eV), along with a ferroelectric $T_C$ > 1000 °C. Another significant benefit of AlScN lies in its capability to be deposited at a low temperature 400 °C using physical vapor deposition (PVD). This characteristic renders it an appealing contender for nonvolatile memory (NVM) that is compatible with BEOL processes. Moreover, the $P_R$ of $Al_{1-x}Sc_xN$ remains remarkably consistent across a range of thickness spanning from 400 nm down to 5 nm.[119,135,152,177,185] This attribute presents a viable strategy for effectively scaling down dimensions in ferroelectric memory applications, addressing the challenges of miniaturization. $Al_{1-x}Sc_xN$ thin films having a thickness of ~ 400 nm demonstrated structural stability up to 1025 °C.[186] Stable $P_R$ with minimal changes were observed up to 400 °C in an $Al_{0.7}Sc_{0.3}N$ thin film on a Pt (111) bottom electrode.[135] The substantial and sustained remanent polarization at elevated temperatures (> 400 °C) coupled with the thermal durability of $Al_{1-x}Sc_xN$ up to 1100 °C positions it as a promising avenue for data storage in computing applications operating in demanding environments (> 500 °C). The ferroelectric memory characteristics utilizing AlScN/GaN heterostructures are reported to operate at 400 °C with the max On-Off ratio of ~10.[123] $Al_{0.68}Sc_{0.32}N$ ferroelectric diode based NVM devices that can reliably operate with clear ferroelectric switching up to 600 °C and distinguishable On and Off states were recently reported. The devices exhibit high remnant polarizations (> 100 $\mu C/cm^2$) which are stable at high temperatures up to 600 °C. At 600 °C, these devices show 1 million read cycles without failure and stable On-Off ratio above 1 for > 60 hours.[119] The ferroelectric diode is technically a resistive memory device (see section c below) that switches in resistance as a function of ferroelectric polarization switching and has a non-destructive read as oppose to FeRAM discussed above for perovskite oxide ferroelectrics which is capacitive in nature and has a destructive read.

### c. Resistive memory

Broadly speaking, Resistive random-access memory (RRAM) technologies are BEOL compatible with CMOS processes, exhibit modest switching energy, fast read and write speeds, and high contrast in resistance states.[187,188] The vast majority of RRAM technologies rely on redox chemistry within the channel or at the electrode interface to modulate the resistance state of the memory element.[189] For example, in filamentary RRAM, oxygen vacancies in transition metal oxides (TMO) sandwiched between metal electrodes form or break conductive filaments in response to specific voltage pulses. As such, memory retention in these devices is limited by the activation energy associated with defect formation and ion migration, even in the absence of any electric field. Recent computational studies[190-192] indicate that local temperature gradients play a significant role in the filament formation in TMOs. Filament formation dynamics change significantly above 125° C, reducing both the reliability and endurance of the memory states.[192] Experimental studies have confirmed that write endurance degrades rapidly above 100° C due to spontaneous dissolution of the conductive filaments.[188,193] For example, the expected retention of a trilayer $Al_2O_3/HfO_2/Al_2O_3$ memory is 10 years at room temperature but only 1 year at 85° C.[194] Similarly, conductive pathways form or break via conductive bridges (CBRAM) during set/reset pulses on metal ions dissolved in a solid electrolyte. Because the change in resistance is associated with metal ion migration, the same activation energy constraints are present in CBRAM as in the filamentary oxide RRAM, and similar limitations on operational temperatures are observed.[195,196] While most amorphous oxide CBRAMs have limitations going to temperatures > 100 °C, recently, 2D heterostructures made out of graphene/$MoS_{2-x}O_x$/graphene have shown CBRAM scalable up to 340 °C with an ON/OFF < 10 up to $10^5$ seconds.[137]

Beyond CBRAM, non-filamentary RRAM such as GaO$_x$ based devices with ITO electrodes have also been demonstrated recently operating up to 327 °C with ON/OFF < 5 for at least 5000 seconds, operating on a mechanism of space charge limited current under high injection.[136] Another related switching mechanism for resistive switching has been demonstrated in correlated-electron memory (CeRAM), which has recently shown promise for high temperature operation. In CeRAM (or Mott) devices, the local oxidation state of transition metal atoms switches in response to changes in local carrier density[197] rather than ion or defect motion. The ultimate limitation on operation temperature for CeRAM is governed by thermal excitation of carriers into and out of the defect band, which is 0.96 eV above the valence band in carbon doped nickel oxide (NiO).[198] Memory retention was demonstrated up to 24 h at 200 °C[198], and single switching operation was demonstrated at 300 °C in carbon doped NiO CeRAM.[199]

Using the concept of electrochemical redox reactions in solid-oxide fuel cells at elevated temperatures, electrochemical RAM (EC-RAM), another class of RRAM capable of operating at elevated temperatures has also been recently demonstrated. These devices operate on the principle of programmable ionic conductivity of oxides as a function of oxygen concentration. Unlike the above discussed RRAM devices that are strictly two terminals, these devices typically have three terminals akin to a transistor where the "gate" injects or withdraws oxygen ions from the channel oxide leading to the change in resistivity between the source and drain. Typical oxides used in these devices for elevated temperatures are transition metal oxides such as WOx, TiOx and rare earth ternary oxides as channel materials while hafnia, Cr$_2$O$_3$ and yttria stabilized zirconia (YSZ) are used as electrolyte materials.[200-202] The best high T NVM performance has been shown in WO$_{3-x}$ devices with YSZ electrolyte gate where a retention of 10$^5$ secs for ON/OFF > 1 was shown at 200 °C.[200] Thus far, ECRAMs have shown excellent analog programmability and a kbit level array with ~ 140 °C operation has also been shown.[201] However, programming speeds, ON/OFF ratios and extended retention at temperature remain a concern. Based on the operating principle of controlling oxygen ion concentration, oxide based ECRAMs are unlikely to survive at temperatures > 500 °C as oxygen exchange between environment and between materials becomes prevalent.[202]

4. **Metals for contacts and interconnects:**

Elevated temperature operation requires very specific properties and conditions from metal interconnects and contacts. The most important features are reliability and integrity of the electrical connections for prolonged exposure to the temperature of operation and the operating electrical load.[14,30] In this regard, thermal and chemical stability together with resistance to electromigration and surface diffusion at temperature become critical. Therefore, metals with high melting points (refractory metals) and strong barrier to oxidation (noble metals or barrier oxide forming metals) are often preferred over the most conductive metals for interconnects/contacts in semiconductor devices operating at elevated temperatures. In addition, minimal chemical reactivity/alloy formation tendency with the semiconductor and a close matching of CTE with the semiconductor are also important (Figure 4) to minimize differentials in thermal stress which might damage the interface or deleteriously change operational parameters. Given these criteria, combined with the electronic criteria of work-function necessary for many devices discussed above, the set of metals used in high-T devices are narrowed down to a handful of options. Among them refractory elemental metals like W, Mo, Hf, Zr as well as noble, non-corrosive metals such as Au, Pt, Ni and their combinations (bilayer or trilayer stacks) have been widely used.[203-207] Among compound metals, refractory nitrides such as TiN, TaN, MoN, HfN, ZrN [208] and intermetallic silicides such as $TaSi_2$, $TiSi_2$, $WSi_2$, $MoSi_2$, $NiSi_2$, and $PtSi$ have also been used in various cases.[230,231] As seen in Figure 4, CTE-wise, Mo, W and Pd are best matched with most semiconductors and dielectrics for encapsulation/packaging while Ti, Pt, TiN and $TaSi_2$ also exhibit reasonable CTE mismatch. However, Au and Ni, both widely used in research case devices, have much higher CTE and strong dependence of CTE with T making them more vulnerable to failure.

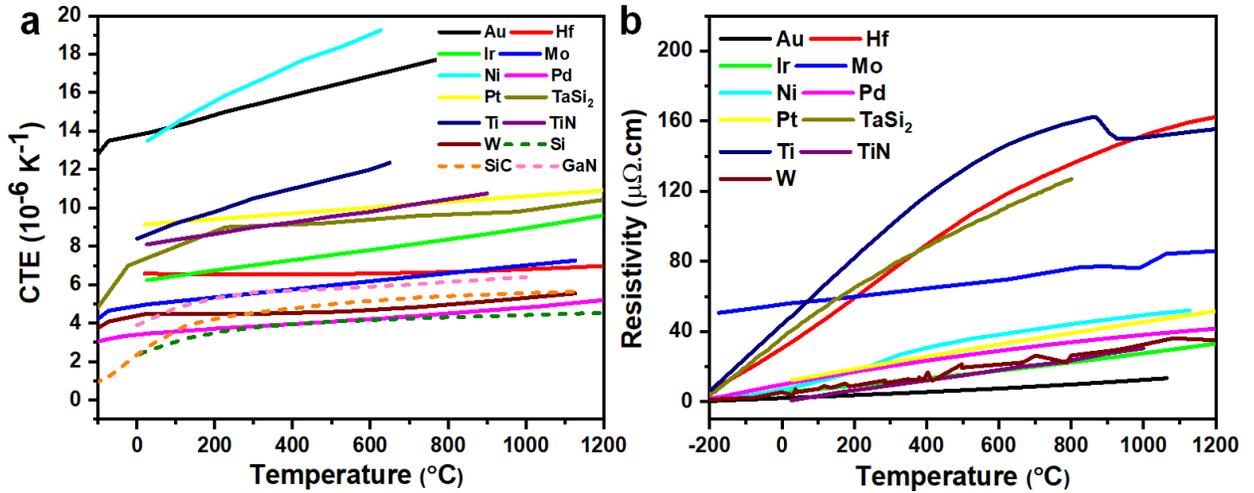

**Figure 4. Materials for contacts and interconnects. a**. Coefficient of thermal expansion (CTE) vs. T for various contact and interconnect metals. Notably W, Mo and Pd have the lowest and flattest dispersion of CTE vs. T. In addition, their CTE values are well matched with the CTE values of wide band gap semiconductors.[209-219] (The dotted lines indicate that the materials are WBG semiconductor).)**b**. Resistivity vs. T for selected metals.[212-217,219-229] It is worth nothing that while Au and Ni have low resistivity and small slope of resistivity vs. T, their relatively high CTE makes then less than ideal. On the other hand, W, Pd and Mo concurrently have low CTE mismatch with the semicondcutors and relatively low resistivity as well as small slopes vs. T making them suitable metals for contacts and interconnects in high T electronics.

For SOI and SiC devices, silicides are the most prevalent contacts and self-aligned silicide formation processes via rapid annealing (silicide process) are well known.[55,232] This is normally accomplished by evaporating the contact metal e.g. Ta, Ti or Ni on the Si or SiC wafer followed by rapid thermal annealing to form the low-resistance $TaSi_2$, $TiSi_2$ or $NiSi_2$ contact at the interface. The melting point of the metal silicides ranges from 1000-2300 °C, and resistivity ranges from 10-120 $\mu\Omega$ cm,[233] therefore they serve as good contacts for Si and SiC high T devices for both JFET and MOSFET type devices. For SiC JFETs, a contact metal stack of Ti/$TaSi_2$/Pt has been well established as stable high temperature contact while a bilayer stack of $TaSi_2$/Pt is a well-established interconnect.[49] For most nitrides and β-$Ga_2O_3$ a combination of Ti/Au, Ti/Ni, Ti/Al/Ni, Ti/Al/Ni/Au, Ni/Au or W have been used as source, drain or gate contacts. There is limited work on interconnects for high T nitride technology with Ti/Au exhibiting stability under relevant operating conditions.[70] For diamond devices elemental refractory metals are often the choice as they form metal carbides upon annealing and form ohmic contacts.[206,207]

To avoid failure at elevated temperatures, barrier layers are equally important on contacts. Nitrides often serve as good barrier layers to prevent diffusion of interconnect metals that lead to their reaction with the semiconductor or overlying metal contacts. TiN and TaN have both served as diffusion barriers in Si microelectronics while TiN has also been shown to be an effective barrier between W and SiC.[234]

## 5. Packaging and integration considerations

Packaging is fundamentally the method by which useful electrical signals are transmitted from the integrated circuit (IC) to the outside environment. The package may contain other components such as passive devices or other ICs which assist with the process of transmitting, modifying, or otherwise performing necessary functions to these electrical signals. Packaging involves trade spaces where often things such as performance and complexity are traded for lifetime, reliability, or any other special requirements for a given application. Therefore, it is difficult to define a best solution or best method for packaging as it inherently involves tradeoffs, particularly in high-T and harsh environment electronics.

Modern heterogeneous packaging can cover a broad range of final products, from a single IC breakout to designs that contain hundreds of passive components integrated with many ICs. The higher the temperature requirement, the less available material options exist for packaging. Most processable plastics soften beyond usability by their glass transition temperature, the best of which are generally below 370 °C[235] with higher temperature polymers being hard to synthesize, process, and largely remain hypothetical.[236] Even polymer composites are generally unsuitable as electronic packaging at extreme temperatures due to their physical[237] or electrical[238] limitations. Solders and interconnects (commonly lead and copper based) begin to oxidize or soften and deform under stress at temperatures as low as 200 °C.[239,240] The lack of industry standards by which to grade the efficacy of any packaging solution in this temperature regime further complicates the qualification of materials. Applications influence what can be used: what may be acceptable for monitoring of a geothermal well may not be acceptable for use in a molten salt nuclear reactor, or for the surface of Venus. This makes comparing one solution to another impractical, as exploration of the packaging system requires destructive analysis and through material characterization.

The operating temperature ranges of common packages and material critical temperatures overlaid with the operating temperature ranges described in this survey as shown in Figure 5a. This chart is not exhaustive and is intended to simplify evaluation of the packaging and materials. For example, to check temperature compatibility for a package at 600 °C, select the point on the blue line for that temperature. Note the materials that fail below this temperature and above this temperature and consider the compatibility of the materials which remain. This figure charts the discovery of a viable semiconductor, substrate and material system that all must converge along the blue line. The solid green box is the operating temperature range of FR4, the most commonly used circuit board material currently employed. This green box also signifies the temperature range which is currently covered by industrial standards, and illustrates how small that range is when compared to the described range of high-temperature electronics.

### a. Ceramic Package Materials

Historically, ceramic packaging has been primarily developed for radio frequency (RF) applications, requiring printed dielectrics and metals to be fired or otherwise cured after patterning to remove solvents and organic binders. Two classes are generally used, co-fired and post-fired ceramics. Co-fired systems involve processing and laminating un-fired cast ceramic tape layers, achieving simultaneous burnout and densification of the ceramic substrate, the conducting vias, and metal traces. Subsequent fabrication steps are referred to as post-fired material processing. Post-fired materials can be built on a substrate that was formed via cofiring, or it can be built up starting from a blank, fully dense, ceramic substrate. To ensure adequate adhesion, co-fired materials require the metals and dielectrics must be shrinkage-matched to the ceramic both in all dimensions, whereas post-fired materials ideally have zero in-plane shrinkage, with the material only shrinking in the z-direction after the firing process has been completed.

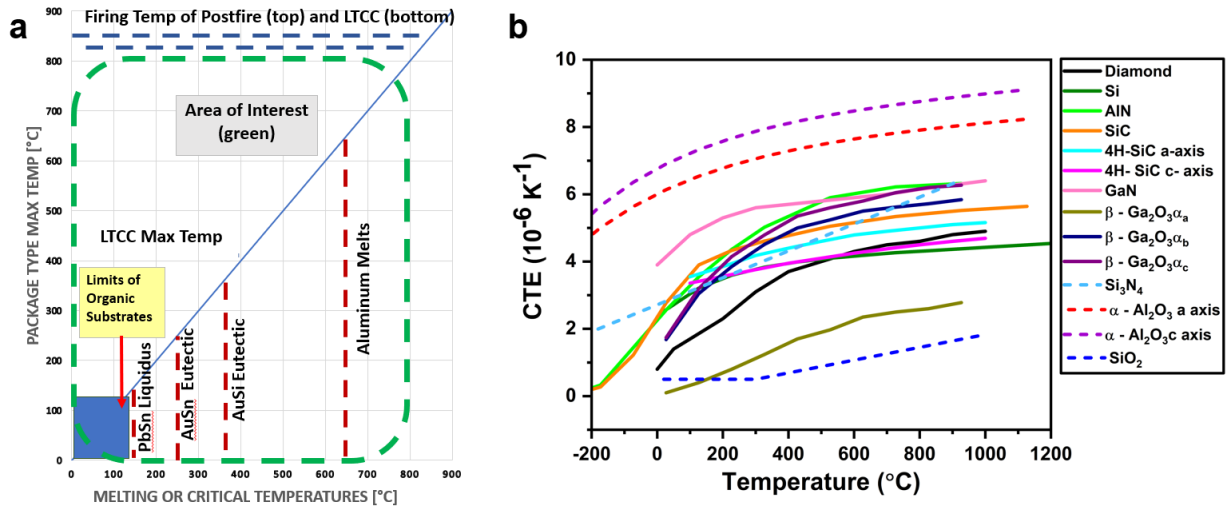

**Figure 5. Materials for packaging and substrates.** a. The horizontal axis is temperature for the materials used in packaging, and the vertical axis represents the temperature for the various package types used for high temperature integration. The blue line indicates the exclusion zone in which materials and packages begin to become unusable due to temperature effects. b. Co-efficient of thermal expansion (CTE) vs T for various semiconductors, substrate ceramics and encapsulation materials.[219,241-261] While CTE varies mildly with temperature for most materials, there is large CTE mismatch between sapphire and most wide band gap (WBG) semiconductors, suggesting SiC as the substrate of choice given the availability of 200 mm wafers. Among encapsulating materials, $Si_3N_4$ possesses minimal mismatch with WBG semiconductors(The dotted lines indicate that the materials are insulator).

Post-fired materials systems can offer higher reliability at temperature at the expense of integration density. As the extreme high-T application space begins to drive innovation in materials science and materials engineering, these RF-matured processes have found new applications for high-temperature high-reliability packaging. The critical material properties now become the minimization of leakage currents, and temperature dependence of the physical and electrical characteristics of the ceramic system. There are challenges when qualifying these systems for use in high-T applications, as there is not a well-established industry standard for testing at extreme temperatures, and manufacturers rarely extract or measure relevant metrics (DC leakage currents, etc.) in the temperature ranges expected for operation. These ceramics, in general, have good Coefficient of Thermal Expansion (CTE) matching to wide bandgap semiconductors such as GaN and 4H-SiC. (Figure 5b). These packages are detailed further in Box 3.

> **Box 3: Basics of Ceramic Packaging Materials**
>
> There are two main categories for cofired ceramic technology: Low-Temperature Cofired Ceramics (LTCC) and High-Temperature Cofired Ceramics (HTCC). HTCC materials are selected for positive CTE such as alumina ($Al_2O_3$) and aluminum nitride (AlN) and that fire at temperatures between 1650-1850 °C, while LTCC materials are based on alumina ($Al_2O_3$) and glass frits to reduce the firing temperature to approximately 900 °C.[262] The cofired ceramic process involves punching or drilling of via cavities in unfired "green" tape layers, filling the vias with a conductive ink, and then screen printing of the metal traces. Once all the layers have been processed, they are then laminated and fired. LTCC can be fired in an atmospheric furnace while HTCC requires the presence of hydrogen gas during firing, which substantially increases the complexity of the firing process. Additionally, each system has different metals available to form the vias and surface metal layers. The metals used for HTCC must survive the high firing temperature of the ceramic and typically include refractory metals such as molybdenum and tungsten. These metals have higher resistance than the metals which can be used for LTCC, such as silver, gold, and palladium alloys. These metals must be carefully selected, based on the components required to be integrated, to ensure that no harmful interactions will occur at the operating temperature of the package. An example of a harmful intermetallic compound (IMC) that may be encountered is the AuAl IMC formed when gold and aluminum are in contact. This interface most commonly occurs when active devices with AlCu bond pads are bonded out to a package with gold or gold-plated leads and can present a major limitation on the lifetime of the package at elevated temperatures.[263]
>
> Postfired ceramic materials are processed by starting with a fired ceramic substrate, unlike cofired materials where the substrate is formed layer by layer. To form the wiring board and vias, conductive, dielectric, and via materials are printed onto the substrate and fired in sequence. This kind of processing can be more time intensive, as more firing cycles are required, but allows for more metallization and bulk ceramic choices. The integration density of postfired systems is lower than cofired systems, as the diameter of the vias has a practical limit when they are formed through screen printing processes. Post-fired ceramic processing enables the use of very high-quality ceramic substrates for long lifetime (>1000 hours), low leakage applications at temperatures at and above 470 °C.[264] As the processing is carried out on fully dense ceramic, Ceramic Wiring Boards (CWB) fabricated in this manner have excellent planarity. Most postfire material systems additionally support printed thick film resistors, reducing the number of discrete components.

### b. Bonding Materials

There are two main groups of bonding materials: conductive and non-conductive. This is compounded by the three main processes involving bonding: component attachment, die attachment, and hermetic sealing. There is no single bonding process that services the complete range of temperatures or applications covered in this survey. The optimal selection of attachment processes is constrained by the requirements of the application – for example, the selection for a jet engine may be optimized for vibration and thermal cycling, whereas that for a Venus lander may be optimized for chemical resistance and long lifetime with no cycling.

Just as with selection of materials in the ceramic packaging process, the bonding materials used must be compatible at elevated temperatures. At this time, there are no industry standards for qualifying processes in the temperature ranges of 500-800 °C, but MIL-STD-883 can be used as a starting point to qualify component attachment, die attachment, and wire bonding processes at room temperature which can then be tested for lifetime at operating temperature.

**i. Wire Bonding**

Wire bonding from integrated circuit to substrate remains a practical method to form the electrical connection from the die to the package and is most often performed with either gold or aluminum bond wire. Aluminum, gold, and copper are the most readily available bond wire materials. Platinum bond wire is available, but wire bonding with platinum is not commonly performed for IC packaging. At higher temperatures, copper and aluminum are not practical due to electromigration effects as well as corrosion and melting. Copper wire bonding forms problematic IMCs with most high-temperature materials under operating conditions both increasing contact resistance and reducing the reliability of the bond.[265,266] As aluminum has a melting point of approximately 660 °C, it is not a good choice for extreme high-temperature applications. Gold wire bonding to gold bond pads is extremely reliable over the entire usable temperature range of gold. Gold wire can be purchased in various purities and alloys, 4N bond wire (99.99% Au) and 2N (99% Au), and Pd-doped 2N (99% Au). Pd-doped 2N Au bonding wire is preferred for high-temperature applications, as the palladium accumulates and forms a barrier at metal interfaces, improving the aging characteristics of the wire bond at temperature.[267] Gold wire bonds can be formed in a wedge or ball-wedge configuration, requiring a separate die attachment process. Alternatively, gold studs can be formed on the die surface and the die is then attached to the substrate with a "flip-chip" thermosonic process. Flip-chip die attach simultaneously achieves mechanical and electrical attachment of the die to the package. In all cases, careful selection of the metals and barrier layers in the pad structure of the active device is required to prevent IMC formation and the ingress of oxygen or other contaminants to the active device. The IRIS (IRidium Interfacial Stack) pad technology developed by NASA has demonstrated reliability at temperatures up to 800 °C,[268] with 1000 hours test at 500 °C.[48,53], using a combination of CTE matching, diffusion barriers, and wire-bondable materials. IRIS also demonstrates the effectiveness of iridium and platinum as valuable interfacial layers to consider when designing bond pads or other diffusion barriers.[269]

**ii. Die and Component Attachment**

Component attach processing forms a conductive connection between discrete electrical components and the substrate. For component attachment processes, the lowest temperature options are gold-eutectic solders and preforms. These materials require gold interfaces on both joints of the bond and are available as gold-tin (eutectic 280 °C), gold-germanium (liquidus 356 °C), and gold-indium (liquidus 485 °C). For higher temperature conductive bonding materials, silver brazing materials can be used, such as silver-copper-indium (liquidus 730 °C) and silver-copper (eutectic 780 °C). AgCuTi crosses the 800 °C barrier with a eutectic temperature of 845-860 °C. For high-purity metals, 99.99% Ag has a eutectic at 962 °C and 99.99% Au has a eutectic at 1064 °C.[270]

Nanoparticle-based conductive bonding materials potentially offer advantages over eutectic or brazing processes, with low attachment temperatures and high temperature tolerances. The nanoparticles that form the material will fuse at a temperature much lower than the melting point of the bulk material, and once the fusion process is complete the joint will have mechanical and electrical properties similar to the bulk material. These nanoparticles based conductive bonding materials vary, with silver being one the more mature material systems, popularized for its use in attachment of SiC power devices where the high strength and low processing temperatures are advantageous.

Of note, many of the eutectic and brazing materials form hermetic seals and are commonly used for that purpose when properly employed in a design.[271-273]

Die attachment processing can be either conductive or insulating, depending on whether backside-of-die (chip) electrical connection is required, but conductive attachment methods can be stronger and more reliable. Conductive attachments to metallized boards and die form IMCs during brazing or processing.[274] The intermixing of the layers of material (die, attachment material, substrate metallization) can offer advantages over temperature when compared to nonconductive adhesion methods which rely on other adhesion mechanisms. NASA GRC has conducted tests for 1400 hours in simulated Venus atmospheric conditions using a conductive Pt-based die attachment material[275]. In general, conductive die attachment is preferred if the die has been metallized with materials which are compatible with the conductive die attachment materials above. Non-metallized die may prove difficult to attach with any method other than flip-chip in the 800 °C regime. The Young's modulus of the bonding material should likewise be considered. CTE mismatches can be mitigated by materials that have lower Young's modulus, and high modulus materials such as ceramics can cause excessive buildup of stresses during thermal cycling or at extreme high temperatures. Ceramic-based attachment materials exist, but the relatively high (>100GPa) Young's modulus[239] of these materials can cause thermal stress buildup.

For extreme high-temperature applications, sealing can be accomplished using the materials aforementioned, including ceramics and metals combined with brazes and attachment compounds. Similarly, for vibration and environmental hardness, encapsulation may be desirable. But as in all endeavors at HT, CTE must be carefully matched, and development of materials dedicated for these tasks remain areas of research and development. [274]

## 6. Manufacturing considerations and perspective for future research directions:

Decades of research and development have led to the prolific advances in materials and processes outlined within this work. Operation to temperatures approaching 1000 °C are now possible. Scaling these technologies to production is a large part of the remaining challenge. The modern semiconductor ecosystem (from base materials to semiconductors and package) was built not just on silicon temperature ranges, but with silicon economics. Beginning with the 10 µm devices in the 1970s (Figure 6) silicon has raced at the speed of Moore's law, doubling the number of transistors per unit area every two years – with a subsequent improvement in minimum device size and performance. Mapping the state-of-the-art in high temperature logic devices in terms of feature size, it becomes obvious that high-T technologies tend to lag silicon – on a logarithmic scale.

Making devices smaller with higher performance is costly and requires high volume production. Such production was easy to achieve for silicon due to its prolific and widespread use. For the near to intermediate future, high-T electronic applications will be limited to aerospace, avionics, automotive, oil/gas exploration and nuclear power. The volume of chips required to meet the worldwide demand for these industries is therefore expected to be much smaller than high performance silicon microelectronics and therefore rapid progress in miniaturization or performance is not expected to happen at the same rate as silicon.

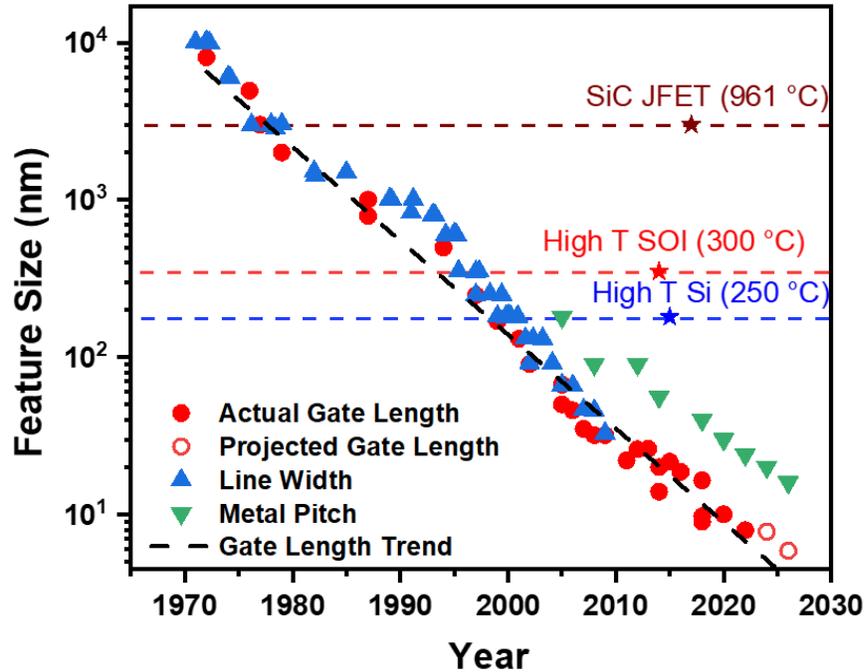

**Figure 6. Scaling and manufacturing outlook.** Moore's law and High-T digital electronics technology in comparison. A tradeoff in node/feature size and highest operation temperature for the various technologies is evident.[8-11,13,42,46-49,53,56,276-286]

This challenge can be overcome, but there is not a single solution. Manufacturing technologies continue to advance. Innovation in direct write (maskless) technologies can be employed today for prototyping, and even small-scale production without significant costs.[287,288] Likewise, packaging and integration can drive down cost. For example, in digital computing at high-T a NVM technology remains elusive and much desired.[119] Integration of such a technology directly within semiconductor thin-films adds cost, however using dense packaging such devices can be demonstrated co-packaged adjacent to a logic microprocessor or even vertically integrated via bump bonding as intermediate steps.

Likewise, testing and evaluation is also expected to contribute significantly to production costs. Meeting qualifications for high-T operations requires advanced and rigorous testing for materials, devices, and system level degradation under various high-T operating conditions, including those combined with corrosive, high-pressure, or high-radiation environments. For the temperatures, there are several variables in testing aside from peak temperature and duration of operation that also vary with the application. These include temperature ramp rates and single vs. multi ramp temperature cycling etc.

To facilitate the development of high temperature electronic components, efforts to standardize manufacturing and testing are required. Throughout this review, we have highlighted key measures of high temperature device performance including retention, endurance, survivability, and packageability. Often due to the nature of scientific inquiry, cutting edge research focuses on pushing a single boundary, but researchers must remain cognizant of the requirements for device deployment and real-world environmental considerations such as temperature ramp rates and cycling. Many of the temperature "limits" reported are from the highest temperature "hero" results, with most device characterization performed at a lower temperature. Additionally, as electrical properties change with temperature, robust characterization over a range of temperatures should be demonstrated, giving insight into the physics of the device as well as reasonable speculation as to if it is intrinsically or experimentally limited.

Standardization of testing protocols and goals should also be revised where the extreme environments force compromises on performance. The industry standard for memory retention is 10 years, which is both untestable and unachievable for most extreme environment technologies. Thus, accelerated test protocols need to be standardized in order to create a solid benchmark between emerging technologies. Often, ideal electronic performance may need to be sacrificed for better survivability via CTE matching. Consequently, when possible, a more practical-oriented approach will yield more useful communication. Establishing a functional threshold such as a reasonably detectable noise limit and then pushing the device to its maximum temperature where it can still operate above that threshold is generally much more useful than achieving a very high on-off ratio or low noise at a pre-defined temperature target.

In conclusion, a thorough account of materials challenges for high-T electronics has been presented across the digital materials landscape including the semiconductor, memory, contact and interconnect metals, dielectrics for insulation and packaging, die attachment and wirebonding. It is clear that as new materials are discovered and developed, a holistic approach is necessary in considering the prospective temperature limitations. For instance, as many of the wide bandgap semiconductors have intrinsic limits beyond 500°C, many times it is the packaging, interconnects, contacts, dielectrics and interfaces that drive the temperature limits for implementation. While the challenges are numerous, as high T electronics technology matures, it will expand from SOI and SiC based technology to other WBG semiconductors which is expected to open up even more applications and opportunities for development of new materials.

**Acknowledgements:** D.J., R.H.O. and D.K.P. acknowledge primary support for this work from AFRL via the FAST programme. D.J. also acknowledges support from the Air Force Office of Scientific Research (AFOSR) GHz-THz program FA9550-23-1-0391. N.G. and D.M. gratefully acknowledge support from Air Force Office of Scientific Research (AFOSR) GHz-THz program grant number FA9550-24RYCOR011.

**Competing Interests Statement:** The authors declare no competing interests.